\documentclass[twocolumn,showpacs,preprintnumbers,amsmath,amssymb,aps]{revtex4-1}
\usepackage{graphicx}
\usepackage{dcolumn}
\usepackage{bm}

\begin{document}

\title{Ultrafast dynamics of electron-phonon coupling in a metal}

\author{Choongyu Hwang$^{1,2}$}\email{ckhwang@pusan.ac.kr}
\author{Wentao Zhang$^{2,3}$}
\author{Koshi Kurashima$^{4}$}
\author{Robert Kaindl$^{2}$}
\author{Tadashi Adachi$^{4,5}$}
\author{Yoji Koike$^{4}$}
\author{Alessandra Lanzara$^{2,6}$}

\affiliation{$^1$ Department of Physics, Pusan National University, Busan 46241, South Korea}
\affiliation{$^2$ Materials Sciences Division, Lawrence Berkeley National Laboratory, Berkeley, CA 94720, USA}
\affiliation{$^3$ School of Physics and Astronomy, Shanghai Jiao Tong University, Shanghai 200240, China}
\affiliation{$^4$ Department of Applied Physics, Tohoku University, Sendai 980-8579, Japan}
\affiliation{$^5$ Department of Engineering and Applied Sciences, Sophia University, Tokyo 102-8554, Japan}
\affiliation{$^6$ Department of Physics, University of California, Berkeley, CA 94720, USA}


\begin{abstract}
In the past decade, the advent of time-resolved spectroscopic tools has provided a new ground to explore fundamental interactions in solids and to disentangle degrees of freedom whose coupling leads to broad structures in the frequency domain. Time- and angle-resolved photoemission spectroscopy (tr-ARPES) has been utilized to directly study the relaxation dynamics of a metal in the presence of electron-phonon coupling. The effect of photo-excitations on the real and imaginary part of the self-energy as well as the time scale associated with different recombination processes are discussed. In contrast to a theoretical model, the phonon energy does not set a clear scale governing quasiparticle dynamics, which is also different from the results observed in a superconducting material. These results point to the need for a more complete theoretical framework to understand  electron-phonon interaction in a photo-excited state. 
\end{abstract}

\maketitle

\section{Introduction}
In the past decade, the advancement of laser technology has allowed extension of spectroscopic tools into the time domain. This has opened a new range of opportunities, especially when it comes to the study of complex materials where more than one interaction is at play and degrees of freedom are often strongly intertwined. This has lead to an unprecedented access to fundamental elementary interactions in materials at their natural timescales.  Among them major emphasis has been dedicated both theoretically and experimentally to the study of electron-phonon coupling that drives new ground states such as superconductivity, Peierls transition etc. 

Angle-resolved photoemission spectroscopy (ARPES) is the ideal tool to study electron-phonon interaction, as it can directly probe the quasiparticles dispersion, Fermi velocity, and self-energy renormalization. Moreover, its ability to resolve such interactions in a momentum space provides unique information on the anisotropy of such interaction. Given the recent extension of ARPES to the tine domain, indeed, electron-phonon coupling can be measured and disentangled from other energy/time scales by directly measuring the relaxation channels of hot electronic distribution in a transient state~\cite{David,Mauri,Wentao}. 

The electron-phonon coupling is also a limiting factor of intrinsic mobility of charge carriers via ultrafast energy dissipation~\cite{Millis,Allen,Chen}. Especially in metals, the electron-phonon coupling leads to a faster relaxation of excited electrons through emitting phonons, when the energy of the excited electrons are higher than phonon energy.  This relatively slower and faster dynamics across the phonon energy generates an energy cut-off in the transient spectra, also known as cut-off energy, and observed in high temperature superconductors~\cite{Jeff,Chris}. However, most of the experimental works so far have focused on the relaxation dynamics of superconductors. In order to have a full understanding and be able to disentangle other interactions, we first need to learn how electron-phonon interaction, in general, responds to photo-excitations. 

\begin{figure*}
\includegraphics[width=2\columnwidth]{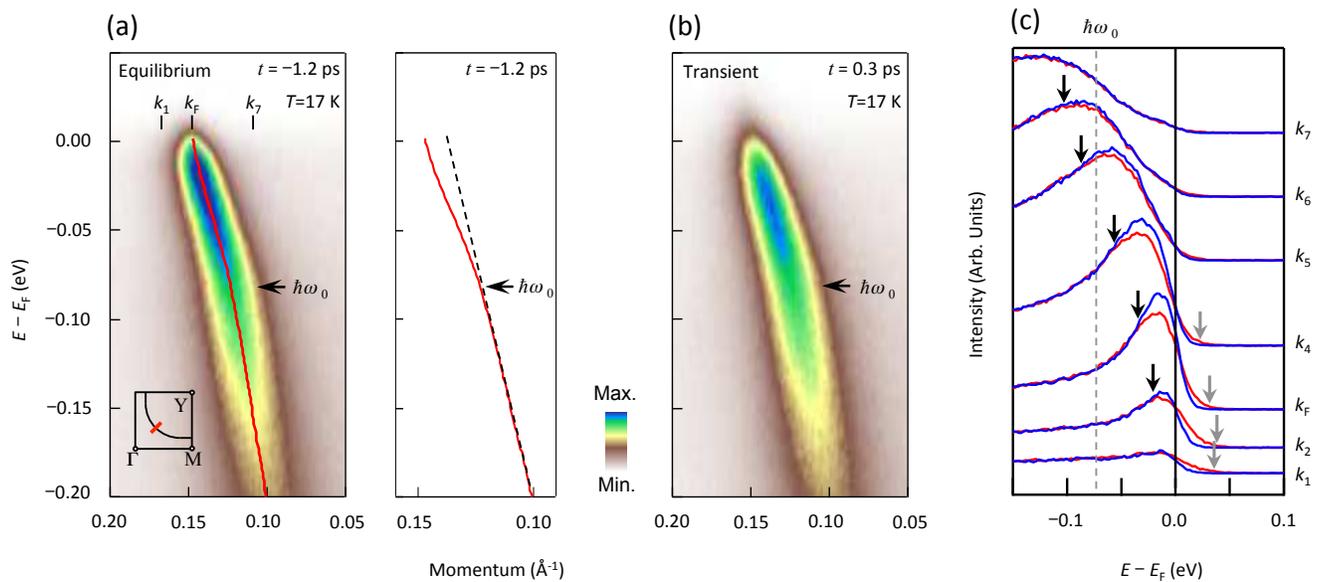}
\caption{(a-b) Equilibrium (a) and transient (b) ARPES intensity maps taken at $t=-1.2$~ps and $t=0.3$~ps, respectively. Both energy maps were taken along the diagonal direction of the Brillouin zone denoted by the red line in the inset of panel (a). Pump fluence is 30~$\mu$J/cm$^2$. The red curve is the Lorentzian fit to momentum distribution curves taken from the ARPES intensity map measured at $t=-1.2$~ps and the black dashed line is a guide to the eyes. Phonon energy is denoted by $\hbar\omega_0$. (c) Energy distribution curves for equilibrium (blue) and transient (red) maps taken at different momentum values from $k_1$ to $k_7$ denoted in panel (a).}
\label{fig1}
\end{figure*}

Here, a metallic system has been examined using time-resolved ARPES (tr-ARPES). Due to an optical excitation, electrons are injected in the unoccupied states, transferring spectral weight toward lower energy states via emitting phonons.  In contrast to the predicted behavior~\cite{Devereaux,Miller} and experimental observation in cuprates superconductors~\cite{Jeff,Wentao}, we find that, in the metallic state, relaxation process gradually increases as the excited carrier density increases (e.\,g.\,, by increasing pump fluence), without a clear energy cut-off that distinguishes slow versus fast dynamics across the phonon energy. These results point to the need for a more complete theoretical framework to understand  electron-phonon interaction in a photo-excited state as well as more detailed work to understand the role that such cut-off energy can play for superconductivity~\cite{Miller}.

\section{Methods}
The time-resolved ARPES measurements were performed at 17~K on heavily overdoped Bi$_{1.76}$Pb$_{0.35}$Sr$_{1.89}$CuO$_{6+\delta}$, which is known to be in a metallic phase. The laser system is a cavity-dumped, mode-locked Ti:sapphire oscillator (Coherent Mira) generating $\approx$150~fs pulses at 840~nm at a repetition rate of 54.3/n~MHz (n=1, 2, 3,...), whose details are described elsewhere~\cite{Jeff_detail}. A transient state is created with an infrared laser pump pulse with an energy of 1.48~eV and measured via photoemission process as a function of delay time with a resolution of 300~fs using an ultraviolet probe pulse with an energy of 5.9~eV. The energy and momentum resolutions are 22~meV and 0.003~\AA$^{-1}$, respectively. The pump fluence is varied from 4 to 24~$\mu$J/cm$^2$.

\section{Results and Discussion}

Figures~\ref{fig1}(a) and (b) show equilibrium and transient ARPES intensity maps at $t=-1.2$~ps and $t=0.3$~ps, respectively. Here $t=0$ represents the time when pump and probe pulses are applied to the sample simultaneously. Negative time refers to equilibrium spectra, e.\,g.\,, spectra taken before the arrival of the pump pulse, while positive time refers to transient spectra, e.\,g.\,, after the sample has been excited by the pump pulse. 

\begin{figure*}
\includegraphics[width=2\columnwidth]{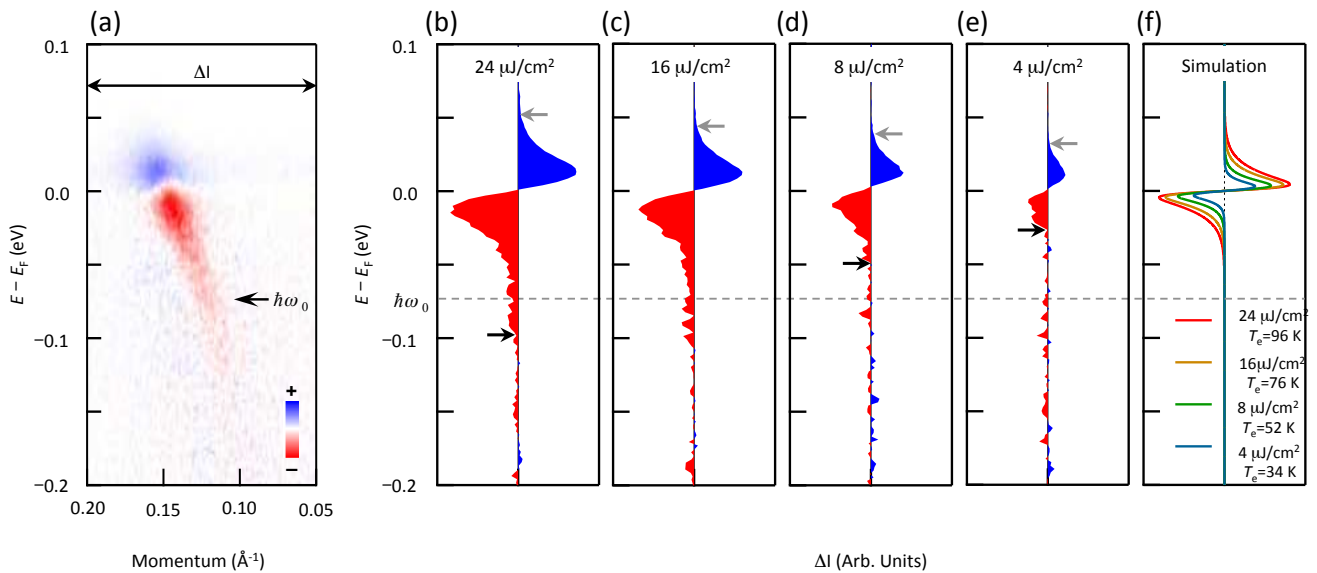}
\caption{(a) Intensity difference map between before and after applying laser pump pulse shown in Figs.~1(a) and~(b), respectively. (b-e) Momentum-integrated spectra ($\Delta$I), from 0.05~${\rm \AA}^{-1}$ to 0.20~${\rm \AA}^{-1}$ in panel (a), for four different pump fluences. (f) Simulation of $\Delta$I assuming the transient heating effect by the pump pulse. Electronic temperature $T_{\rm e}$ at $t=$0.3~ps is 96, 76, 52, and 34~K for a pump fluence of 24, 16, 8, and 4 $\mu$J/cm$^2$, respectively.}
\label{fig2}
\end{figure*}

The data were taken along the Brillouin zone diagonal denoted by the red line in the inset of Fig.~\ref{fig1}(a), using a pump fluence of 24~$\mu$J/cm$^2$. The sample temperature in the equilibrium state was $T=17$~K, measured using a diode thermally connected to the sample. In both maps, one can identify a characteristic energy $\hbar\omega_0\sim70$~meV (denoted by an arrow) that separates a well-defined and poorly-defined dispersive features above and below it, respectively, with different slopes (the black curve is a Lorentzian fit to momentum distribution curves and the black dashed line is an arbitrary straight line for the guide to the eyes). The change in the slope of the energy spectra, so-called a kink, is a universal feature of cuprates due to electron-phonon coupling, especially in the absence of competing phases against superconductivity in the heavily overdoped regime~\cite{Lanzara}. The main effect of the pump pulse is to change spectral intensity. Figure~\ref{fig1}(c) shows energy distribution curves at momentum positions from $k_{\rm 1}$ to $k_{\rm 7}$, denoted in Fig.~\ref{fig1}(a), for equilibrium (blue curve) and transient (red curve) maps. The spectra are normalized by the intensity at higher energy ($-$0.20~eV$\leq$$E-E_{\rm F}$$\leq$$-$0.18~eV) for comparison.

This change in the transient map is clearly evidenced when its photoelectron intensity is subtracted by that of the equilibrium map as shown in Fig.~2(a): loss (gain) in spectral intensity is denoted by red (blue) color. Figures~2(b)-(e) show integrated intensity ($\Delta$I) over a momentum range from 0.05 to 0.20~\AA$^{-1}$. At 24~$\mu$J/cm$^2$, most of the intensity change is observed near $E_{\rm F}$ as shown in Fig.~2(b). However, the intensity change persists beyond $\hbar\omega_0$ down to $E-E_{\rm F}\sim-$100~meV (denoted by a black arrow). Upon decreasing pump fluence, the cut-off energy is also decreasing to $\sim-$50~meV at 8~$\mu$J/cm$^2$ as shown in Fig.~2(d) and $\sim-$30~meV at 4~$\mu$J/cm$^2$ as shown in Fig.~2(e). Overall decrease of $\Delta$I is due to the decreasing pump fluence that excites less electrons to a transient state. Additionally, the cut-off energy above $E_{\rm F}$ is far below $\hbar\omega_0$ at 24~$\mu$J/cm$^2$ as denoted by a gray arrow in Fig.~2(b) and approaches toward $E_{\rm F}$ with decreasing pump fluence. Such a spectral change cannot be explained by thermal smearing of the Fermi edge due to the transient heating effect by the pump pulse, because the observed cut-off energy is higher than the cut-off energy expected from electronic temperature ($T_{\rm e}$) increased by the pump pulse, e.\,g.\,, Fig.~2(f) shows the difference of Fermi-Dirac distribution at $T_{\rm e}$ from that at the equilibrium temperature, 17 K. These observations are in contrast to a proposed model where the intensity cut-off originates from the electron-boson coupling~\cite{Devereaux}.

Such coupling provides a fast decay channel for electron (hole) excitations above (below) the boson energy. High energy excitations can decay by emitting (absorbing) a boson. Boson-mediated decay is predicted to reduce the life time of quasiparticles to tens of femtoseconds. Along this line, the efficacy of this additional bypass is determined by the electron-boson coupling strength, which can be measured by the kink strength. However, the kink strength of Bi$_{1.76}$Pb$_{0.35}$Sr$_{1.89}$CuO$_{6+\delta}$ are roughly similar across different doping and temperature. This apparent inconsistency can be explained by the difference between single particle life time (which can be extracted from the self-energies) and population life time (the quantity ARPES is measuring)~\cite{Yang}. The difference exists when an energy conserving decay channel, for instance, Coulomb and elastic impurity scattering, is present~\cite{Yang,Rameau}. When a single particle decays through such a channel where energy is retained in the electronic system, another excitation is created. As the process cascades, the population takes much longer to vanish after than the single particle time scale.

This intensity change is accompanied by an intriguing time-dependence of electron self-energy as shown in  Fig.~3. Within the Debye model, both real and imaginary parts of electron self-energies are affected by electron-phonon coupling~\cite{Hengsberger,Verga,Reinert}. In other words, electrons acquire renormalized mass as well as enhanced scattering through the coupling to a phonon~\cite{Verga}. Indeed, real part of electron self-energy, Re$\Sigma$, shown in Fig.~3(a), obtained by subtracting a bare or unrenormalized band (assumed to be a linear band) from the measured or renormalized energy spectrum shown in Figs.~1(a) and~(b), exhibits a peak-like shape around $\hbar\omega_0$ due to the electron-phonon coupling, consistent with previous reports on cuprates~\cite{Lanzara,Verga,Reinert}. Upon applying laser pump pulse at 24~$\mu$J/cm$^2$, Re$\Sigma$ decreases from blue filled circle at $t=-1.2$~ps to red filled circle at $t=0.3$~ps, when the difference, $\Delta$Re$\Sigma$, is denoted by gray shaded area. Re$\Sigma$ is recovered back to the almost original shape after longer delay time at $t=10.2$~ps, denoted by blue empty circles. Hence, the electron-phonon coupling is suppressed upon applying laser pump pulse by $\sim$14~\%, determined by the change of slope of Re$\Sigma$ near $E_{\rm F}$, with a line fit to the data from $-$50~meV to $-$20~meV shown as black lines, and recovered afterward. Similar self-energy change is observed when the pump pulse is incident on a a superconducting cuprate~\cite{Wentao}. However, a critical difference is that for a non-superconducting cuprate, overall self-energy is depressed by the pump pulse even below the phonon energy $\hbar\omega_0$. On the other hand, for a superconducting cuprate, self-energy change is prominent mostly for a self-energy peak at $\hbar\omega_0$, whereas for the others self-energy change is barely observed.

\begin{figure}[t]
\includegraphics[width=1\columnwidth]{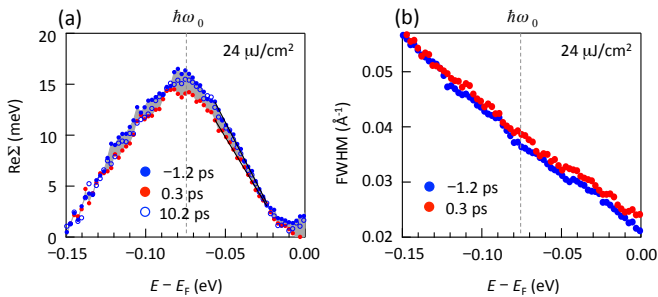}
\caption{(a) Real part of electron self-energy (Re$\Sigma$) before ($t=-1.2$~ps) and after ($t=0.3$~ps and 10.2~ps) applying laser pump pulse. (b) Full width at half maximum (FWHM) of the energy spectra before ($t=-1.2$~ps) and after ($t=0.3$~ps and 10.2~ps) applying laser pump pulse with three difference fluences.}
\label{fig3}
\end{figure}

Figure~3(b) shows the full width at half maximum (FWHM) of the momentum distribution curves taken as a function of $E-E_{\rm F}$ for the measured energy spectra shown in Figs.~1(a) and~(b). FWHM is considered to be linearly proportional to the imaginary part of electron self-energy (Im$\Sigma$) by FWHM=2$\vert$Im$\Sigma$/{\it v}$\vert$, where {\it v} is the bare velocity. Upon applying laser pump pulse at 24~$\mu$J/cm$^2$, FWHM at $E_{\rm F}$ is enhanced by $\sim$14~\%, indicating the suppression of the electron-phonon coupling. With decreasing $E-E_{\rm F}$, however, the different gradually decreasing, until the difference becomes negligible at $E-E_{\rm F}\sim0.12$~eV. \
{Consistent with the Re$\Sigma$, while the biggest changes occurs at lower binding energy, the pump induced changes are also reflected above the phonon energy.} The change in Re$\Sigma$ and FWHM indicates that electronic excitation suppresses the strength of electron-phonon coupling in a transient state with respect to an equilibrium state and that the suppression is recovered as a function of time. This hinders to extract the strength of electron-phonon coupling in an equilibrium state from the dynamics of electrons in a transient state, in contrast to the claim of the previous theoretical work on a non-superconducting metal~\cite{Devereaux}.

\begin{figure}[t]
\includegraphics[width=1\columnwidth]{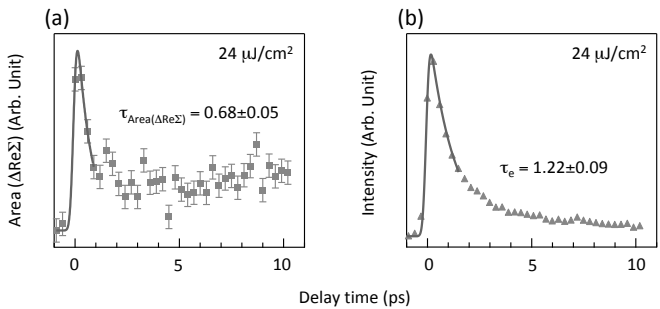}
\caption{(a) The area of the difference in electron self-energy as a function of delay time. Solid curve is an exponential least-square fit to extract the time constant of 0.68$\pm$0.05~ps. (b) Averaged spectral intensity above $E_{\rm F}$ as a function of delay time. Solid curve is an exponential least-square fit to extract the time constant of 1.22$\pm$0.09~ps.}
\label{fig4}
\end{figure}


To further understand the time-dependent electron-phonon coupling, the dynamics of $\Delta$Re$\Sigma$ has been studied as a function of delay time and compared to the quasiparticle dynamics as extracted from the integrated intensity above $E_{\rm F}$~\cite{Jeff}. The area of $\Delta$Re$\Sigma$ for an energy range of $-$0.15~eV$\leq$$E-E_{\rm F}$$\leq$$-$0.02~eV shows an initial sharp increase near $t$=0 with a pump fluence of 24~$\mu$J/cm$^2$ that is relaxed with increasing delay time. The solid line in Fig.~\ref{fig4}(a) is an exponential least-square fit to Area($\Delta$Re$\Sigma$) that gives a time constant, $\tau_{{\rm Area}(\Delta{\rm Re}\Sigma)}$, of 0.68$\pm$0.05~ps. Interestingly, the time scale, associated with the kink, is by a factor of two smaller than the relaxation time of quasiparticles: $\tau_{e}$ of 1.22$\pm$0.09~ps as shown in Fig.~4(b). A similar difference between the two time scales has been also observed for cuprates superconductors~\cite{Wentao}. However, $\tau_{{\rm Area}(\Delta{\rm Re}\Sigma)}$ is almost 5 times faster than the one in a superconducting cuprate, e.\,g.\,, 3.12~ps for Bi$_2$Sr$_2$CaCu$_2$O$_{8+\delta}$~\cite{Wentao} versus 0.68~ps for a non-superconducting cuprate observed in Fig.~4(a), at the same pump fluence. In other words, the suppressed coupling between electrons and phonons are recombined even faster in a non-superconducting (and metallic) cuprate than for a superconducting cuprate.

Indeed in a metallic system, the electron population resembles a thermal population that is the case of a non-superconducting cuprate. In contrast, when a gap opens in a superconducting cuprate, electron relaxation dynamics become coupled to the dynamics of the electron population and phase restriction processes kick in, leading to coexisting femtosecond and picosecond dynamics~\cite{ChrisPRB}. Clearly these results suggest that there are several mechanism at play that could provide new insights on the nature of the superconductivity and ask for more detailed theoretical works.

\section{Summary}
We have shown that, in a non-superconducting cuprate, i.\,e.\,, a metallic system, the electron-phonon coupling is perturbed by a femtosecond laser pump pulse in contrast to the theoretical prediction and that the perturbation is relaxed in a faster time scale compared to that in a superconducting cuprate. In addition, the suppressed coupling between electrons and phonons are recombined even faster in a non-superconducting cuprate than for a superconducting cuprate, suggesting a possibility that several mechanism may play an important role in electronic dynamics in the high temperature superconductivity. Our results not only provide the time scale of the coupling formation between electrons and phonons in a non-superconducting system, but also invite a more complete theoretical framework to understand electron-phonon interaction in a photo-excited state.

\acknowledgments
This work was primarily supported by Berkeley Lab’s program on Ultrafast Materials Sciences, funded by the U.S. Department of Energy, Office of Science, Office of Basic Energy Sciences, Materials Sciences and Engineering Division, under contract DE-AC02-05CH11231. CH also acknowledges support from the National Research Foundation of Korea (NRF) grant funded by the Korea government (MSIT) (No. 2017K1A3A7A09016384 and No. 2018R1A2B6004538).

\end{document}